\documentclass[prb,aps,showpacs,twocolumn,unsortedaddress]{revtex4}
\usepackage{graphics,bm}
\usepackage{amssymb}
\usepackage{epsfig}
\usepackage{epsf}
\begin{document}

\title{Electronic Interface Reconstruction at Polar-Nonpolar Mott Insulator Heterojunctions}

\author{Wei-Cheng Lee}
\email{leewc@mail.utexas.edu}

\author{A.H. MacDonald}
\email{macd@physics.utexas.edu}
\affiliation{Department of Physics, The University of Texas at Austin, Austin, TX 78712}

\date{\today}

\begin{abstract}
We report on a theoretical study of the electronic interface reconstruction (EIR) 
induced by polarity discontinuity at a heterojunction between a polar and a nonpolar Mott
insulators, and of the two-dimensional strongly-correlated electron systems (2DSCESs) which accompany
the reconstruction.  We derive an expression for the minimum number of polar layers required
to drive the EIR, and discuss key parameters of the heterojunction system which control 
2DSCES properties.  The role of strong correlations in enhancing confinement at the interface 
is emphasized. 
\end{abstract}

\maketitle

\section {Introduction}
Two-dimensional (2D) electron systems have been a fertile source of interesting physics over the past few decades, 
playing host to the fractional and integer quantum Hall effects and cuprate superconductivity among other phenomena. 
The most widely studied and most thoroughly understood 2D electron systems 
are those that occur near semiconductor heterojunctions.  In these systems 
carrier densities and disorder strengths can be  
adjusted using modulation doping and the electric field effect, and high sample quality 
can be achieved using 
lattice matched materials and epitaxial growth techniques.
These 2D systems are well described by Fermi liquid theory, at least at magnetic field $B=0$.
Rapid recent progress in the epitaxial growth of complex transition-metal 
oxides\cite{ohtomo2, ahn, keimer, fitting, hotta, ohtomo, nakagawa, thiel} foreshadows the 
birth of an entirely new class of 2D electron systems, one in which electronic correlations are strong
even at $B=0$ and non-Fermi-liquid behavior is common.

The present authors have recently argued\cite{lee} that the modulation doping technique commonly used 
in semiconductor heterojunction systems can also be applied to complex-oxide heterojunctions
to create high-quality two-dimensional strongly-correlated electron systems (2DSCESs)
localized near the interface between two different Mott insulators. 
Our study was based on a generalized Hubbard model which captures some key features of these
systems, and on combined insights achieved by applying Hartree-Fock theory (HFT), dynamical mean-field theory
(DMFT) and Thomas-Fermi theory (TFT) approaches.  This theoretical strategy echoed that adopted by Okamoto and Millis\cite{okamoto} 
in addressing the two-dimensional electron systems which can occur near the interface between 
a Mott insulator and a band insulator.  In both cases the properties of the 2D electron systems
reflect a delicate balance between space-charge electric fields and strong local correlations.

These 2D systems 
are clearly unusual in many respects. For instance, their quasiparticles 
interact with the 3D spin-excitations of the Mott insulator 
barrier materials.  Since researchers now appear to be at the cusp of achieving 
experimental control over this kind of 2D electron system, 
it is interesting to explore the possibilities theoretically
in an attempt to find useful theoretical frameworks, and hopefully
also to partially anticipate some of the new phenomena likely to occur. 
To mention one example, it has recently been suggested that a spin-liquid insulator
could be found in a Mott insulators bilayer\cite{rib} system with an 
appropriate doping profile.

The apparent electronic interface reconstruction (EIR) discovered\cite{ohtomo,nakagawa,thiel} at the heterojunction between
the band insulators LaAlO$_3$ and SrTiO$_3$ is unique in complex oxide interface studies and does not have an analog in semiconductor heterojunction systems,
although there is still some uncertainty about the role of oxygen vacancies in the measured conductivity. For example, it has been proposed \cite{pen}
that the oxygen vacancies are responsible for the insulating behavior observed at the $p$-type interface.
The reconstruction is forced by the polarity difference between LaAlO$_3$ and SrTiO$_3$.
Polarity discontinuities are normally weaker at semiconductor heterojunctions and, when 
present, the electric fields to which they give rise are usually
screened by relaxation of atoms near the interface\cite{sze}. In complex transition metal oxides, however,
it was discovered\cite{Sawatzky,nakagawa} that the polarity discontinuity field can be screened electronically
by transferring electrons between surface and interface layers, changing the valence of transition metal ions in
these layers.  Electrons dodge the {\it polar catastrophe}\cite{nakagawa} without essential atomic assistance. 
It has been shown\cite{thiel} that such an interface can be tuned between metallic and
insulating states by relatively modest electric fields, suggesting the possibility of novel
electronic devices.

In this paper we examine EIR at a heterostructure between model polar and nonpolar Mott insulators. 
We have in mind for these effects Mott insulators with cubic peroskite structures.  Possible materials combinations are discussed in Section VI.
The long-ranged Coulombic space-charge fields 
are treated within the Hartree approximation, and strong on-site Coulomb interactions
are treated by solving HFT, DMFT, and TFT equations. We find that 
a minimum number of polar layers is required to induce an EIR because of strong local correlations, whereas
the number of nonpolar layers does not play an essential role in determining the electron distribution.
We also find that when a 2D electron system is present near the heterojunction, it is strongly localized
because of strong correlations which limit the potential of the electronic system to lower its
energy by spreading out in the growth direction.
TFT results provide a simple way of understanding the more microscopic DMFT and HFT 
calculations and are remarkably accurate.
 
\begin{figure}
\includegraphics{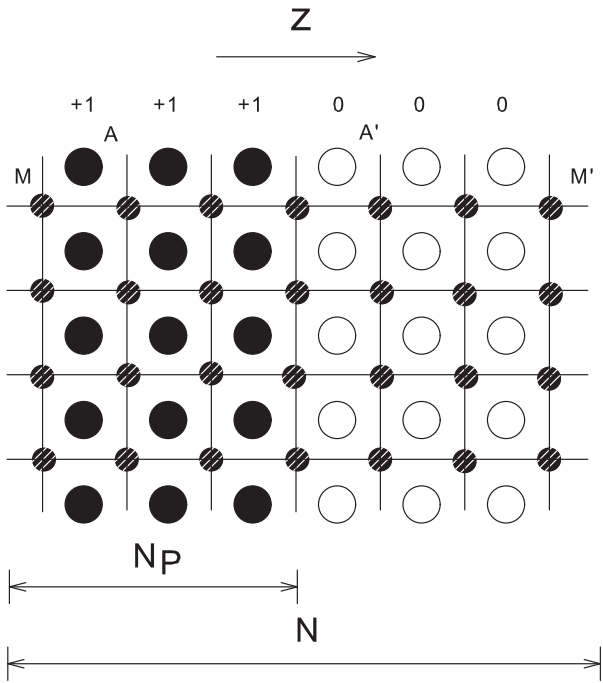}
\caption{\label{fig:one} Schematic illustration of our model for a thin film containing a polar-nonpolar Mott insulator heterostructure.  The model consists of $N_p$ layers of a polar Mott insulator perovskite ($AMO_3$) and $N-N_p$ layers of 
a nonpolar Mott insulator perovskite ($A'M'O_3$). The $\hat{z}$ direction
is chosen to be the layer-by-layer growth direction and the charge density is assumed to be uniform in each $x-y$ plane. The symbol $A$ represents a group III element with nominal 
valence $+3$ in the $AO$ layer while the symbol $A'$ represents a group II element with
nominal valence $+2$ in the $A'O$ plane.  The $AO$ layer therefore has a surface charge density of $+1$ per $A$ atom while the 
$A'O$ layer is neutral.  The symbols $M$ and $M'$ represent transition metal ions in group IV and V respectively. Both $M$ and $M'$ have $3d^1$ configuration before the electronic interface reconstruction (EIR).}
\end{figure}
\section {Single Band Hubbard Model}
Fig. \ref{fig:one} illustrates the model heterostructure we investigate in this paper. 
We consider a thin film composed of $N_p$ polar perovskite Mott insulator layers ($AMO_3$) and $N-N_p$ 
nonpolar perovskite Mott insulator layers
($A'M'O_3$), where $A$ ($A'$) is a group III (II) element with a valence of +3 (+2) in the $AO (A'O')$ layer. 
$M$ ($M'$) is a group $IV$ ($V$) element which is a $3d^1$ Mott insulator because of strong local 
repulsion among the d-orbitals.  In this study we disregard the interesting complications associated with orbital 
degeneracy and do not account directly for hybridization between the transition metal and oxygen ions. 
We therefore use a  single band Hubbard model to describe the $d$ valence electrons. 
Long-ranged Coulomb interactions are described realistically by accounting for charges at the oxygen sites and 
on the $A$, $A'$, and $M$, and $M'$ ions. The resulting model Hamiltonian is:
\begin{equation}
\begin{array}{l}
\displaystyle 
H=H_d + H_U + H_{Coul}\\[2mm]
\displaystyle
H_d = \sum_{i\sigma} \epsilon_{d}(z_i) \; d^\dagger_{i\sigma} d_{i\sigma} -\sum_{<i,j>\sigma} t_{ij} (d^\dagger_{i\sigma} d_{j\sigma} + h.c.)\\[3mm]
\displaystyle
H_U=\sum_i \; U(z_i) \; \hat{n}_{i\uparrow}\hat{n}_{i\downarrow}\\[3mm]
\displaystyle
H_{Coul}=\frac{1}{2}\sum_{i\neq j,\sigma,\sigma'}
\frac{e^2 \; \hat{n}_{i\sigma}\hat{n}_{j\sigma'}}{\epsilon \vert\vec{R}_i
-\vec{R}_j\vert}\\[2mm]
\displaystyle
\,\,\,\,-\sum_{i,\sigma}\sum_{j(j_z\leq N_p)}\frac{Z^A e^2 \; \hat{n}_{i\sigma}}
{\epsilon\vert\vec{R}_i-\vec{R}^A_j\vert} -
\sum_{i,\sigma}\sum_{j(j_z>N_p)}\,\frac{e^2 \; \hat{n}_{i\sigma}}
{\epsilon\vert\vec{R}_i-\vec{R}_j\vert}
\end{array}
\label{eq:hamiltonian}
\end{equation}
where $i$ labels a metal site and (grouping coplanar oxygen and A cation charges)
$Z^A=1$ in the present case. 
The d-orbital hopping term is expressed in  
Eq.(\ref{eq:hamiltonian}) as a sum over links; most of our calculations have been performed for a 
model with nearest-neighbor hopping.  Second nearest-neighbor hopping has little influence on our main results,
but will impact the 2DSCES Fermi surfaces as discussed below.  The sizes and alignments of the gaps in the two 
$3d^{1}$ Mott insulators are fixed by the interaction and site energy parameters:
$U(z_i)=U_P$ and $\epsilon_{d}(z_i) = \epsilon_{d1}$ for the polar layers and 
$U(z_i)=U_{NP}$ and $\epsilon_{d}(z_i) = \epsilon_{d2}$ for the nonpolar layers. 
In an ideal cubic perovskite unit cell with lattice constant $a$, 
$\vec{R}_i=a(n_i,m_i,z_i)$ and $\vec{R}^{A,A'}_i=a(n_i+1/2,m_i+1/2,z_i+1/2)$ respectively. 
The third term in $H_{Coul}$ accounts for a net effective charge per $M'$ atom in the $M'O_2$ plane; this charge
is the sum of the oxygen ion charges and the $+5$ charge of the $M'$ ion when in a $d^0$ configuration.  In this work 
we have neglected the uneven distribution of charge within each layer.
In this calculation we for the most part assume that 
$\epsilon_{d}$ is constant, but emphasize that the difference
between d-orbital energies on opposite sides of the heterojunction can be an 
important parameter.  As in case of semiconductors, band lineups play a key 
role in determining heterojunction properties and are likely to be difficult to 
estimate with sufficient precision.  Accurate values will most likely have to 
be determined experimentally for each heterojunction system considered. 
The requirement of overall electrical neutrality in the thin
film implies that the d orbital occupancy, averaged over the thin film $\bar{\rho}$, satisfies 
$\bar{\rho} = 1$.

Throughout this study we use the Hartree approximation for the long-ranged Coulomb interaction and 
neglect the corresponding exchange term, since our goals in studying this toy model are purely qualitative. 
The self-consistently determined Hartree mean-field Hamiltonian is
\begin{equation}
H^{eff}_{Coul}=\sum_{i\neq j,\sigma}
\frac{e^2 (\rho_j-f)\hat{n}_{i\sigma}}{\epsilon \vert\vec{R}_i
-\vec{R}_j\vert} -
\sum_{i,\sigma}\sum_{j(j_z\leq N_p)}\frac{Z_A e^2 \hat{n}_{i\sigma}}{\epsilon\vert\vec{R}_i
-\vec{R}^A_j\vert}
\end{equation}
where $\rho_j=\sum_\sigma\langle\hat{n}_{j\sigma}\rangle$ is the electron
density on site $j$, $f=0$ for $j_z\leq N_p$, and $f=1$ for $j_z> N_p$. 
In the continuum limit, $H^{eff}_{Coul}$ becomes:
\begin{equation}
\frac{H^{eff}_{Coul}}{2\pi\,U_c}=\sum_{i\sigma}\left[\sum_{z_A}\vert i_z-z_A\vert - \sum_{z'\neq z}(\rho(z')-f)\vert z'-i_z\vert\right]\hat{n}_{i\sigma}
\end{equation}
where $U_c=e^2/\epsilon a$.

\section {Hartree-Fock Theory}
In HFT the Hubbard $U$ term is replaced by its mean-field counterpart: $n_{i\uparrow}
n_{i\downarrow}\to \sum_\sigma \langle n_{i-\sigma}\rangle\,n_{i\sigma}$. As pointed out 
\cite{okamoto,lee} previously, there are almost always a number of distinct self-consistent solutions of the HF equations,
corresponding to various local-minima and saddle-points of the HF energy functional.
Here we present results for solutions with $G$-type antiferromagnetic (AFM) order, or $(\pi,\pi,\pi)$ spin order, and ferromagnetic (FM) order.
These solutions are representative of ordered states which occur frequently in complex transition metal 
oxides and are often global minima of the HF energy functional.  The philosophy of following this procedure is 
that HFT can neither reliably judge the 
competition between different types of order, nor predict whether or not long-range order succumbs to fluctuations.
The results nevertheless very often provides a reasonable description of local correlations in the strongly
correlated regime.

\begin{figure}
\includegraphics{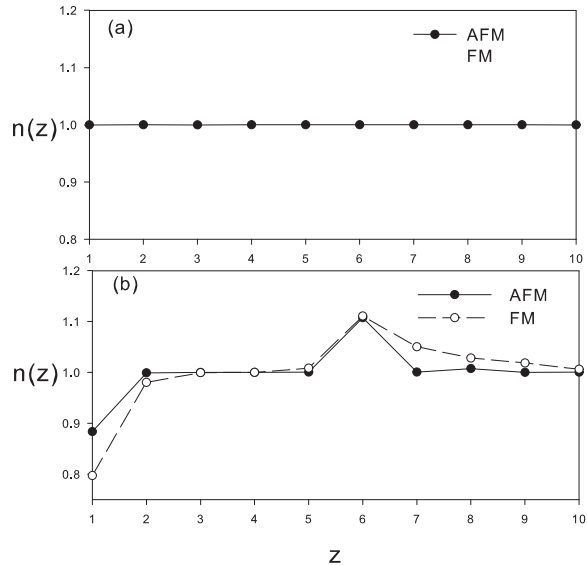}
\caption{\label{fig:two} Electron charge distributions for AFM and FM 
Hartree-Fock states with $U_P/t=U_{NP}/t=20$, $U_c/t=0.8$, and $\epsilon_{d}$ constant
for (a) $N_p=2$ and (b) $N_p=5$ cases. The total number of layers is $N=10$.
EIR occurs only in the model thin film with the larger number of polar layers. 
Confinement of the 2DSCES is stronger in the AFM state solution 
because of its stronger local correlations.}
\end{figure}

\begin{figure}
\includegraphics{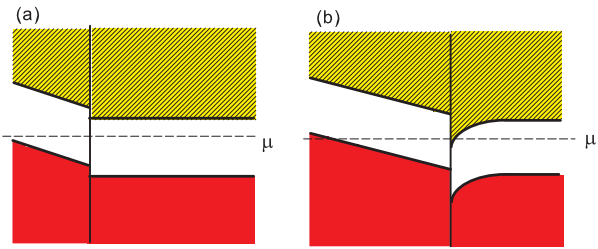}
\caption{\label{fig:three} (Color online) Cartoon illustration of band bending for (a) $N_p=2$ (b) $N_p=5$. The support of the upper Hubbard band's
spectral weight is indicated by yellow shading while the support of the 
lower Hubbard band's spectral weight is indicated by solid red.  These illustrations reflect 
the spectral weight only at the $M$ and $M'$ transition metal sites of 
$M(M')O_2$ layers with the end point on the polar side of the interface at
the last $MO_2$ layer and the starting point on the nonpolar side of the interface at the first 
$M'O_2$ layer. Since the last $MO_2$ and the first $M'O_2$ layers are 
separated by a $AO$ layer, the electrostatic potential reaches its minimum at the first $M'O_2$ 
layer. (The electrostatic potential is of course continuous along the growth direction, as required by the Poisson equation.)
(a) $\mu$ lies in the gap region and no EIR occurs. (b) $\mu$ crosses the lower Hubbard band 
on the polar side and upper Hubbard band on the nonpolar side, indicating the appearance of EIR.
If the top layer of the polar material was a positively charged $AO$ layer rather than a negatively 
charged $MO_2$ layer, the charge transfer would occur from the lower Hubbard band near the heterojunction to the 
upper Hubbard band near the surface.}
\end{figure}

Our HFT results for $N_p=2$ and $N_p=5$ are summarized in Fig. \ref{fig:two}. 
For the calculations reported here, the total layer number $N=10$ and typical values 
were chosen for the interaction parameters.
($U_P/t=U_{NP}/t=20$ and $U_c/t=0.8$.)\cite{okamoto}
For both AFM and FM states EIR occurs only for the model thin film with
the larger value of $N_p$.  This feature is a result of the competition between the polar catastrophe\cite{nakagawa}
and strong local correlations.  Although the polarity discontinuity at the interface favors
a transfer of electrons from the outmost surface to the 
interface, this charge rearrangement is opposed by the Mott-Hubbard gap.
It is instructive to consider space-charge induced band bending 
diagrams like the cartoon illustration of Fig. \ref{fig:three} which contrasts
the $N_p=2$ and $N_p=5$ cases. For $N_p=2$, the bending of the upper and lower Hubbard bands 
is not large enough to force the chemical potential $\mu$ at the heterojunction outside
the gap region. When more polar layers are grown, $\mu$ eventually crosses the lower Hubbard band of the 
outermost polar layer and the upper Hubbard band of the nonpolar layer nearest the interface, forcing the occurrence of an EIR.
If the top layer of the polar material was a positively charged $AO$ layer rather than a negatively 
charged $MO_2$ layer, the charge transfer would occur from the lower Hubbard band near the heterojunction to the 
upper Hubbard band near the surface.  

\begin{figure}
\includegraphics{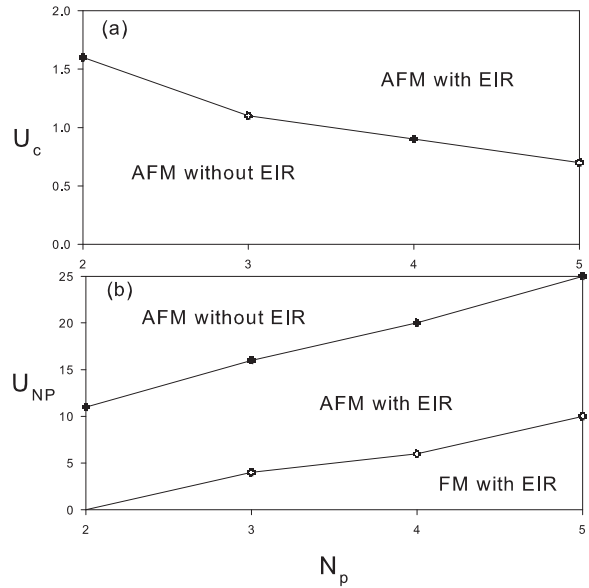}
\caption{\label{fig:four} Ground-state phase diagram (a) {\em vs.} $N_p$ and $U_c/t$ for  $U_P=U_{NP}=20$ and (b) 
{\em vs.} $N_p$ and $U_{NP}$ for $U_P=20$ and $U_c/t=0.8$. The total number of layers is fixed
at $N=10$. (a) For fixed $U_P$ and $U_{NP}$, EIR occurs for a smaller number of polar layers if $U_c$ is larger.
Note that the FM states always have higher ground-state
energy than AFM states in these parameter regions.
(b) For fixed $U_P$ and $U_c$, EIR requires more number of polar layers when $U_{NP}$ is larger.
As indicated here, a smaller $U_{NP}$ also favors FM states over AFM states in Hartree-Fock theory.  
For both (a) and (b) the phase boundary for the occurrence of EIR is accurately reproduced 
by the Thomas-Fermi approximation, Eq.\ref{nmp}.}
\end{figure}

An analogous competition has already been discussed and 
studied experimentally in LaAlO$_3$/SrTiO$_3$ heterostructures \cite{thiel}. 
In that case the energy which competes with the polar 
catastrophe is the energy difference between the top of oxygen $p$-bands in nonpolar SrTiO$_3$ and the bottom of aluminum $d$ bands
in LaAlO$_3$, instead of the correlation gap in the example studied here. 
Since it always costs the gap energy to add more electrons on any layer of the heterostructure,
electrons accumulate initially near the interface to gain the most electrostatic energy, 
eventually spreading out through the heterostructure to minimize kinetic energy cost.
Quantum confinement of the 2DSCES associated with EIR is stronger for states with stronger on-site correlations because 
the kinetic energy is then a smaller component of the overall energy budget.
This is the reason that in Fig. \ref{fig:two}(b)
electrons are more confined to the interface in the AFM state than in the FM state.
(The FM state has larger bandwidths and smaller correlation gaps, at least within
HFT.) Based on the discussion above, we expect that for fixed $U_P$ and $U_{NP}$, the minimum number of polar 
layers required to achieve reconstruction $N^m_p$ will be smaller when $U_c$ is larger. 
$N^m_p$ will also tend to decrease when the Hubbard gap is reduced in either layer;
in Fig. \ref{fig:four} we illustrate the dependence on $U_{NP}$ when $U_P$ is fixed. 
Note that a smaller value of $U_{NP}$ favors ferromagnetic HF states.

\begin{figure}
\includegraphics{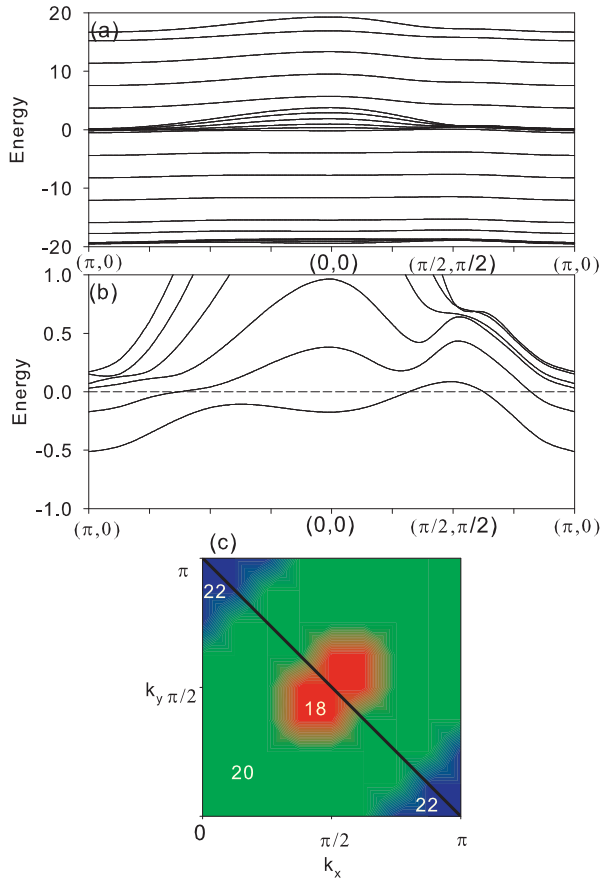}
\caption{\label{fig:five} (Color online) Electronic structure of the $N_p=5$ AFM state 
whose charge distribution was illustrated in Fig.\ref{fig:two}(b).  This plot was calculated 
with a non-zero second nearest-neighbor hopping parameter $t'/t=-0.15$.
Energies are measured from the chemical potential of the thin film.
(a) All 20 2D bands along high symmetry lines of the AFM Brillouin zone. (b) Crossing between the 
polar material surface layer lower Hubbard band and the five non-polar material upper Hubbard band layers.
The space charge physics implies that these six bands will be close to the chemical potential. 
Note that only two bands are partially occupied.  (c) Dependence of the total band occupancy on position in the 
Brillouin-zone on a color scale with occupancy 18 in red, 20 in green, and 22 and in blue.}
\end{figure}

A more microscopic view of EIR is provided in Fig.\ref{fig:five} where we illustrate the electronic structure of the 
AFM solution of the Hartree-Fock equations for a thin film with $N_p=5$.  The parameters used to 
construct this illustration are the same as those used for the $N_p=5$ case in Fig.\ref{fig:two}(b).
The $N=10$ thin film has 20 two-fold spin-degenerate 2D bands because of the reduced translational symmetry 
of the AFM state, and charge neutrality requires that 10 bands be occupied.  In Fig.~\ref{fig:five}(a) 
we can identify 10 relatively widely spaced bands which correspond at high energies to the upper Hubbard band 
and at low energies to the lower Hubbard band in the polar material.  The bands are widely spaced because of the 
average electric field in the polar material; for the same reason these bands are quite highly localized in 
individual atomic layers.  The two groups of more narrowly spaced bands correspond respectively to the upper Hubbard band 
and the lower Hubbard band of the non-polar material.  Because the space-charge electric fields are much either 
almost fully or at least partially screened out in the non-polar material by charge accumulation at the interface,
the bands are closely spaced and the corresponding eigenstates contain more inter-layer characters.  When EIR occurs there is a 
weakly avoided anticrossing between the top-most lower Hubbard band state in the polar material and the upper Hubbard band states of the 
non-polar material.  This anticrossing is apparent in Fig.\ref{fig:five}(b) which expands the region of the electronic structure
close to the Fermi energy.  In the ground state the polar lower-Hubbard band state is lower in energy at the 
Brillouin-zone center but higher in energy toward the Brillouin-zone edges.  (For a model with only nearest-neighbor
hopping the lower upper band has its maximum and the upper Hubbard band has its minimum along the line from $(\pi,0)$ to $(0,\pi)$.) 

For the parameters used in this calculation (with second nearest-neighbor hopping $t'/t=-0.15$) the EIR leads to two partially occupied 2D bands
illustrated in Fig.~\ref{fig:five}(c), a nearly full lower-Hubbard band very localized on the surface and with two 
inequivalent hole pockets in the AFM Brillouin zone centered on $(\pi/2,\pi/2)$ and $(\pi/2,-\pi/2)$ 
and a nearly empty band concentrated on the first non-polar
layer with an electron pocket centered on $(\pi,0)$.   

\section {Thomas-Fermi Theory}
Layered oxide materials have a natural Thomas-Fermi approximation in which the total 
energy is expressed as the sum of the total electrostatic energy and a local-density-approximation
for the band and correlation energies.  For layered structures the contribution to the energy from 
each atomic layer is approximated by the energy per layer of an electrically neutral 
3D system with the average site occupancy of that layer.  Minimizing this energy with a fixed total 
density constraint leads to the following TF equation:
\begin{equation}
\mu(\rho(z))+v_H(z)=\mu_0
\label{eq:TF} 
\end{equation}
where $v_H(z)$ for this case is.
\begin{equation}
\frac{v_H(z)}{2\pi\,U_c}=\sum_{z_A}\vert z-z_A\vert - \sum_{z'\neq z}(\rho(z')-f)\vert z'-z\vert
\end{equation}
and $\mu_0$ is the overall chemical potential. 
Following the same approach described in Ref\cite{lee}, we can solve the TF equations for different 
ordered states using $\mu(\rho)$ calculated from HF solutions for the three-dimensional Hubbard model. 
We confirmed that TFT very accurately reproduces the results of HFT
for the charge distribution between the layers, as in the case of modulation doping\cite{lee}.

One advantage of using this TF equation is that we can derive some key properties analytically.
For example, if we assume that the d-orbitals of polar and non-polar sides are both described 
by one-band Hubbard models and that the charge transfer occurs from a single surface layer to a single interface layer 
Eq.(\ref{eq:TF}) reduces to 
\begin{equation}
\epsilon_{d1} + \mu_P(1-\delta n) + 2 \pi U_c N_{p} (1 - 2\,\delta n) = \epsilon_{d2} + \mu_{NP} (1+\delta n)  
\end{equation}
where the Hubbard model chemical potentials are expressed as function of density in units of electrons per atom,
$\delta n$ is the charge transfer, and the on-site energies which are usually set to zero in Hubbard models, 
restored on each side of the heterojunction.  Since the left hand side is a monotonically decreasing function of 
$\delta n$ and the right hand side is a monotonically increasing function of $\delta n$, 
this equation has a solution and EIR will occur if 
\begin{equation}
2 \pi U_c N_{p} \geq [\epsilon_{d2} + \mu_{NP}(1+)] - [\epsilon_{d1} + \mu_P(1-)] = E_{IG} 
\label{nmp}
\end{equation}
{i.e.} if the electric potential drop across the polar layer exceeds the spatially indirect gap $E_{IG}$ between the bottom of the upper Hubbard band
in the non-polar material and the top of the lower Hubbard band in the polar material.  Since the variation of chemical potential with density 
is strongly reduced outside of the gap in strongly correlated material, it follows that the amount of charge transfered 
once $U_c N_{p}$ exceeds the minimum value is given accurately by 
\begin{equation}
\delta n \simeq \frac{2 \pi U_c N_{p} - E_{IG}}{4 \pi U_c N_{p}}.
\label{eq:deltan} 
\end{equation}
As the right hand side of Eq.~\ref{eq:deltan} becomes comparable to 0.5 at larger values of $N_{p}$, 
the fundamental assumption of our simple model (namely that the polar catastrophe is avoided by electronic rather than atomic 
reconstruction) becomes increasingly suspect.  When EIR occurs it is likely to lead to relatively small
2D Fermi surface pockets, as illustrated in Fig.~\ref{fig:five}.

\section {Dynamical Mean-Field Theory}
\begin{figure}
\includegraphics{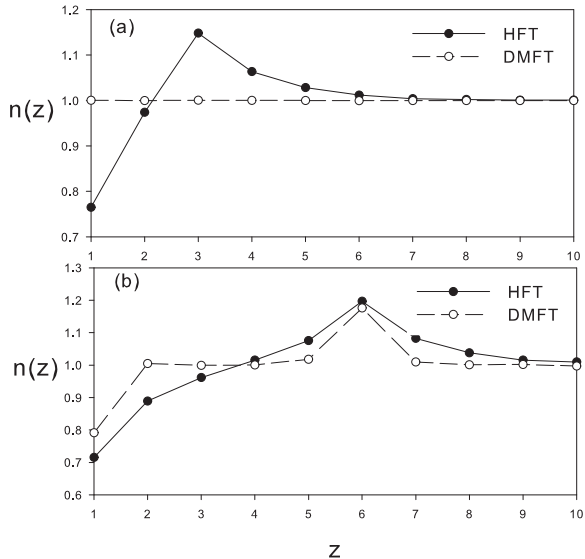}
\caption{\label{fig:six} Electron distribution calculated by HFT and DMFT for PM state with (a) $N_p=2$ and (b) $N_p=5$. Other parameters are the same as those used in
Fig \ref{fig:two}. The comparison between HFT and DMFT
for PM state demonstrates the role of on-site correlation in the EIR. The solutions of DMFT show that EIR only occurs with more polar layers and the confinement is
stronger compared to HF solutions.}
\end{figure}
Hartree-Fock theory can describe the physics of strong local correlations only when the correlations are static and 
lead to FM or AFM ordered states; HF solutions for PM states completely miss local correlation and do not 
reproduce the Mott-Hubbard gap.  DMFT captures correctly the insulating
gap created by on-site correlations even for inhomogeneous multilayered systems\cite{okamoto,lee,freericks}. 
For this reason DMFT calculations for paramagnetic states (PM) provide important additional insight into EIR in Mott-Hubbard insulator 
heterojunction systems.  For the impurity solver necessary in applications of DMFT, we adopt the 
simple two-site method proposed by M. Potthoff\cite{potthoff},
which has been successfully applied to a variety of different transition metal oxide heterostructures\cite{okamoto, lee}.
The results for $N_p=2$ and $N_p=5$ are presented in Fig. \ref{fig:six} with $U_P/t=U_{NP}/t=20$ and $U_c/t=0.8$. 
When a paramagnetic state is described by HFT, charge transfer from the surface 
to the interface occurs (incorrectly) even for $N_p=2$. 
For $N_p=5$, the screening of the surface layer electric field incorrectly begins immediately 
in the polar material. The 
DMFT calculations demonstrate that a larger value of $N_p$ is required even if 
the spin-degree of freedom is not ordered in the Mott insulators. These 
results confirm that a minimum thickness of polar layers is required for EIR when on-site correlations are strong.

\section{ Material Considerations}
It is not so easy to determine on purely theoretical grounds which 
material combinations could be used to realize the EIR proposed here. In the family of cubic perovskites, 
candidate $d^{1}$ polar materials include LaTiO$_3$ and YTiO$_3$ and possible $d^{1}$ 
non-polar materials  include SrVO$_3$ include CaVO$_3$.  The band offsets 
of the various possible materials combinations $\epsilon_{d1}-\epsilon_{d2}$ 
can be estimated from existing electronic structure calculations. 
LDA calculations based on Wannier functions\cite{pavarini} indicate that the $t_{2g}$ bands
tend to be higher in energy relative to the oxygen $p$ levels in LaTiO$_3$ and YTiO$_3$  
than in CaVO$_3$ and SrVO$_3$. 
Assumming that the oxygen $p$ levels are closely aligned across the heterostructure, 
we can expect that LaTiO$_3$ and YTiO$_3$ will 
have higher $d$ orbital energies than CaVO$_3$ and SrVO$_3$.
Another interesting observation from the LDA calculations is that energy differences 
between different $t_{2g}$ orbital  
associated with distortions of the ideal cubic perovskite structure  
tend to increase across the series 
SrVO$_3$-CaVO$_3$-LaTiO$_3$-YTiO$_3$. 
These additional features related to the $t_{2g}$ orbital degree of freedom may bring in new physics which can not be 
explored in our single-band thin film model. Moreover, the tendency toward magnetic ordering may
also be influenced by the presence or absence of $t_{2g}$ orbital 
degeneracy, which will in turn be influenced by reduced symmetry near the heterojunction.
The same calculations show that CaVO$_3$ and SrVO$_3$ are metallic in the bulk.
Lower coordination and electric fields that reduce inter-layer hopping are likely 
to tip the balance toward the insulating state in thin films and especially near 
heterojunctions, so these materials are still candidates for 
realization of the physics studied here.

Although sophisticated first principle calculations
can provide some insight and will help with the construction of realistic phenomenological models,
the consequences of orbital degeneracy or near-degeneracy for complex oxide heterojunctions may depend on 
subtle issues of many-particle physics.  Another aspect not captured realistically in this qualitative study is the dielectric constant which 
in general could be a very complicated function of positions throughout the heterostructures\cite{sawa}. 
Experimental information when available will likely play an essential role in achieving a full understanding.

One of the most important effect of the band lineups in our
model is its influence on where the electrons accumulate when EIR occurs. 
We have shown that in the case of constant $\epsilon_d$
electron accumulation near the heterojunction appears on the first $M'O_2$ layer.
This will change if $\epsilon_d$ for the polar materials is low enough so that
the sum of $\epsilon_d$ and electrostatic potential is minimized at the last $MO_2$ layer, 
instead of the first $M'O_2$ layer.   

\section {Summary}
In this paper we have studied electronic interface reconstruction (EIR) in polar-nonpolar Mott insulator heterostructure systems.
We use a simple one-band Hubbard model supplemented with long-range Coulomb 
interactions which assumes that the polar catastrophe in layered polar materials
is avoided by electronic rather than lattice reconstruction.  We find that EIR occurs only if a thin
film contains a minimum number of polar layers $N^m_p$; when model paramaters representative of 
perovskite oxides are chosen $N^m_p \sim 4$. 
In this paper, the electronic structure of polar-non-polar Mott insulator heterojunctions
was investigated using Hartree-Fock theory (HFT), Thomas-Fermi theory (TFT), and 
dynamical mean-field-theory (DMFT).  Using TFT we are able to derive an analytical
expression for the dependence of EIR on model parameters, which may be useful in selecting 
suitable materials.  For polar and non-polar Hubbard interaction parameters ($U_P$ and $U_{NP}$ respectively)
a stronger Coulomb interaction parameter $U_c$ will reduce $N^m_p$,
while for $U_P$ and $U_c$ fixed, larger $U_{NP}$ will increase $N^m_p$. 
Strong on-site correlations enhance quantum confinement of the interface states which 
accompany EIR, resulting in 2D systems that are nearly completely confined to an 
atomic plane.  When the Mott insulator barrier materials have antiferromagnetic
order, Hartree-Fock calculations demonstrate the emergence to 2D electron systems with 
Fermi surface areas and characters that depend sensitively on the details of the model.  
The quasiparticles of this system will differ from ordinary 2D electron systems because they
will interact strongly with the 3D spin excitations of the antiferromagnetic Mott insulators.
The unusual properties suggest the likelihood of interesting non-Fermi liquid states 
in some regions of the model's parameter space.

We have also discussed some possible material combinations which might lead to the EIR we envision and 
the key role of the alignment of d-orbital energies across the heterojunction. 
Some realistic features of perovskites, for example, orbital degeneracy and lattice distortions, are not 
captured by the simplified model used in this work. 
We note that the exchange term from the long-ranged Coulomb interaction, which we have neglected in this 
study, should enhance the correlation gap. As a result, the charge distribution may be somewhat different if this term is also treated self-consistently.
These features of realistic systems 
are likely to play an important role in practice, especially so when the charge transfer from the 
surface to the interface region is large enough to sensibly alter the chemical bonding
as discussed previously for the Mott-Insulator/Band-Insulator heterojunction case\cite{hamann}.

In our view, it will likely never be practical to attempt a complete and accurate description of complex oxide interfaces of any 
type from first principles.  Insights from {\em ab inito} theory, simplified tight-binding models, and most critically 
from experiment are all likely to play a role in achieving robust, predictive, phenomenological models of 
these systems.  This work is intented as a step toward establish the correct form of useful phenomenologies.

This work was supported by the National Science Foundation under grant DMR-0606489 and by the Welch Foundation.
The authors acknowledge helpful interactions with Harold Hwang, George Sawatzky, and Charles Ahn.

\end{document}